\documentclass[fleqn,usenatbib]{mnras}

\usepackage[T1]{fontenc}
\usepackage{ae,aecompl}

\usepackage{graphicx}	
\usepackage{amsmath}	
\usepackage{amssymb}	
\usepackage{pdflscape}  
\usepackage{multirow}   
\usepackage{dcolumn}    
\usepackage{placeins}   
\usepackage{txfonts}

\newcommand{\aips}[1]{\texttt{#1}}
\newcolumntype{d}[1]{D{.}{.}{#1}}   

\hypersetup{pdfauthor={R.M.Jeffrey et al.},
            pdftitle={Jet speeds and flares in SS 433}}

\title[Jet speeds \& flares in SS~433]{Fast launch speeds in radio flares,
  from a new determination of the intrinsic motions of SS~433's jet bolides}

\author[R. M. Jeffrey \& al.]{
Robert M. Jeffrey$^{1}$,
Katherine M. Blundell$^{1}$,
Sergei A. Trushkin$^{2,3}$, and\newauthor
Amy J. Mioduszewski$^{4}$
\\
$^{1}$University of Oxford, Department of Physics, Keble Road, Oxford, OX1
3RH, U.K.\\
$^{2}$Special Astrophysical Observatory RAS, Karachaevo-Cherkassian Republic,
Nizhnij Arkhyz, 36916, Russian Federation\\
$^{3}$Kazan Federal University, Kazan, 420008, Russian Federation\\
$^{4}$NRAO, P.O. Box 2, Socorro, NM 87801, USA
}

\date{Accepted XXX. Received YYY; in original form ZZZ}

\pubyear{2016}

\begin{document}
\label{firstpage}
\pagerange{\pageref{firstpage}--\pageref{lastpage}}
\maketitle

\begin{abstract}
  We present new high-resolution, multi-epoch, VLBA radio images of the
  Galactic microquasar SS~433. We are able to observe plasma knots in the
  milliarcsecond-scale jets more than $50$ days after their launch. This
  unprecedented baseline in time allows us to determine the bulk launch speed
  of the radio-emitting plasma during a radio flare, using a new method which
  we present here, and which is completely independent of optical
  spectroscopy. We also apply this method to an earlier sequence of 39 short
  daily VLBA observations, which cover a period in which SS~433 moved from
  quiescence into a flare.  In both datasets we find, for the first time at
  radio wavebands, clear evidence that the launch speeds of the
  milliarcsecond-scale jets rise as high as $0.32c$ during flaring
  episodes. By comparing these images of SS~433 with photometric radio
  monitoring from the RATAN telescope, we explore further properties of these
  radio flares.
\end{abstract}




\section{Introduction}\label{sec:intro}

Over almost 40 years, astronomers have continued to be surprised by the
variety of phenomena displayed by the Galactic microquasar SS~433's binary
system and outflows. It is one of the few known microquasars whose
relativistic jets have been persistently resolved both in space and in time,
allowing us to probe their evolution. In arcsecond-scale radio images, the
precessing jets manifest themselves in a distinctive distorted corkscrew shape
(\cite{2004ApJ...616L.159B}), while propagating ballistically away from the
launch point at approximately $8\,\text{mas}\,\text{d}^{-1}$. Using the VLA,
the jet plasma is observable over $\sim 18$ months after launch. At
milliarcsecond-scales, the jet can be resolved, and individual pairs of
antiparallel discrete plasma ejections (often referred to as bolides) can be
tracked for about $30$ days. The jet also appears as distinctive moving
emission lines in optical (\cite{1979ApJ...230L..41M}) and X-ray
(\cite{2002ApJ...564..941M}) spectra.

Both the spectral behaviour and the morphology of the radio emission are fairly
well described by the precessing jet model of \cite{1979Natur.279..701A} and
\cite{1981ApJ...246L.141H}. In this ``kinematic model'', antiparallel jets of
plasma are symmetrically launched at $0.26c$ along a precessing ejection
vector, which traces out a cone in space every $163$ days (see
\cite{2001ApJ...561.1027E} for parameter fits from optical
spectroscopy). Superimposed on this, the jet axis displays small, regular
nutations with periodicity of $6$ days (\cite{1982ApJ...260..780K}).

The kinematic model is not, however, the complete picture. It describes the
average behaviour of the jets, which have shown a remarkable persistance and
stablity over a few decades of optical spectroscopy
(e.g. \cite{1989ApJ...347..448M}, \cite{2001ApJ...561.1027E},
\cite{2005ApJ...622L.129B}, \cite{2007MNRAS.380..263C}). But, both in the
arcsecond-scale radio jets (\cite{2004ApJ...616L.159B}) and in the optical
spectra (\cite{2005ApJ...622L.129B}, \cite{2007A&A...474..903B}), there is
clear evidence that the jet speed varies over timescales as short as a few
days. The distribution of launch speeds peaks at roughly the mean value,
$0.26c$, and spreads beyond the range from $0.2c$ to $0.3c$
(\cite{2005ApJ...622L.129B}).

The microquasar's brightness also displays variability in both optical and
radio wavebands. At times, SS~433 enters an active state, in which its radio
brightness may rise by between 50 and 100 per cent or more. It can remain in
this state for as long as 90 days
(\cite{2004ASPRv..12....1F}). Contemporaneous optical and radio observations
of a flare in 2004 (\cite{2011MNRAS.417.2401B}) revealed the coupled,
multi-wavelength behaviour of the flaring phenomenon, and fit with a picture
in which flares comprise enhanced mass loss through faster winds, faster
launch speeds for the optical jet bolides, and a 2-part radio flare arising
first from the wind, and then from the appearance of jet bolides. In
particular, from archival radio and optical spectroscopic data,
\cite{2011MNRAS.417.2401B} report that faster optical jets always precede radio
flares by a couple of days.

In this paper, we outline a method for the determination of jet speed using
spatially-- and temporally--resolved radio maps. Using VLBA data from 2011-12
(Section 3) and 2003 (Section 4), we give the first unique determination of
the speeds of SS~433's radio bolides on milliarcsecond scales. Finally, we discuss the
relationship between bolide luminosity and speed in the context of flaring
episodes.

\section{Calculating jet bulk speeds}\label{sec:calculations}

\begin{figure}
  \begin{center}
  \includegraphics[width=240.0pt]{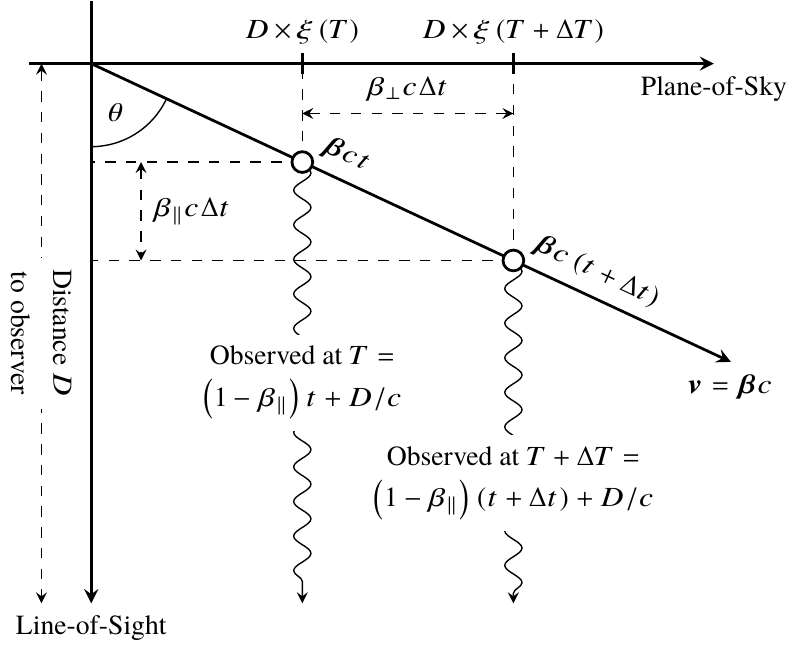}
  \caption{Schematic showing the role of light travel time effects in
    determining proper motions. It shows light pulses leaving a source that
    moves at $\boldsymbol{v} = \boldsymbol{\beta} c$ to be observed by a
    distant ($D \gg \beta c \Delta t$) astronomer on Earth. The events
    corresponding to emission of a light pulse occur at \emph{coordinate}
    times $t$ and $t+\Delta t$ in the observer's rest frame, while the events
    corresponding to those pulses reaching Earth occur at coordinate times $T$
    and $T+\Delta T$ respectively. From the geometry shown here, the time
    steps are related by ${\Delta T / \Delta t =
      \left(1-\beta_\parallel\right)}$, since the second ray has to travel a
    shorter distance from position at emission to observation on Earth. At
    these two observations, the source's angular displacements from the core
    are $\xi\left(T\right)$ and $\xi\left(T+\Delta T\right)$. The proper
    motion measured by the astronomer on Earth is ${\mu =
      \left[\xi\left(T+\Delta T \right) - \xi\left(T\right)\right] / \Delta
      T}$, where $\xi\left(T+ \Delta T\right) - \xi\left(T\right) =
    \beta_\perp c \Delta t / D$.  By eliminating $\Delta t$,
    Equation~\protect\ref{eqn:propermotiondefinition} follows simply.}
  \label{fig:lighttraveltime}
  \end{center}
\end{figure}

\subsection{Proper motions and line-of-sight velocities}
Due to the finite travel time of light signals, the proper motion, $\mu$, of
a light source across the sky is given by:
\begin{align}\label{eqn:propermotiondefinition}
  \mu = \frac{\mathrm{d}\xi}{\mathrm{d}T}
      = \frac{\beta_{\mathrm{apparent}} c}{D} 
      = \frac{c}{D}\frac{\beta \sin{\theta}}{1-\beta\cos{\theta}}
\end{align}
where $D$ is the distance to the source, $c$ is the speed of light, $\theta$
is the angle of the source's intrinsic velocity to line-of-sight, and $v =
\beta c$ is its speed. $\xi$ is the angular displacement on the sky, and $T$
denotes time passed according to the observer's clock (the epoch of
observation) (see Figure~\ref{fig:lighttraveltime} or, c.f.,
\cite{longair2011high} for derivation).

Consider a pair of symmetric, antiparallel, ballistic jets, launched from a
common core, both travelling at speed $\beta$, with the jet and counterjet
aligned at angles $\theta$ and $\pi-\theta$ to the line-of-sight. Inserting
these into the definition of proper motion, we can extract
(e.g. \cite{1994Natur.371...46M}) the line-of-sight velocity component,
$\beta\cos{\theta}$:
\begin{align}\label{eqn:lineofsightvelocity}
  \beta_\parallel   = 
  \beta\cos{\theta} = \frac{\mu_\mathrm{jet} - \mu_\mathrm{cjt}}
                           {\mu_\mathrm{jet} + \mu_\mathrm{cjt}}
                    = \frac{\xi_\mathrm{jet} - \xi_\mathrm{cjt}}
                           {\xi_\mathrm{jet} + \xi_\mathrm{cjt}}
\end{align}
where $\mu_\mathrm{jet}$ and $\mu_\mathrm{cjt}$ denote the proper motions of
the jet and counterjet respectively. The final equality follows from the
assumptions of ballistic motion and simultaneous launch, where
$\xi_\mathrm{jet} = \mu_\mathrm{jet} \Delta T$ and $\xi_\mathrm{cjt} =
\mu_\mathrm{cjt} \Delta T$ are the angular displacements from the launch point
observed at the same epoch (according to the observatory clock), $\Delta T$
after their simultaneous launch.

\subsection{Launch epoch and intrinsic jet speed}
For ballistically moving ejecta, it is now possible to extract both the launch
epoch, $T_\mathrm{launch}$, and the intrinsic speed, $\beta$. To find the
launch epoch, we simply extrapolate back to the date on which the separation
between jet and counterjet ejections was zero. Given two observations at epochs
$T$ and $T+\Delta T$, and defining the total angular separation
$\xi_\mathrm{tot}\left(T\right) := \xi_\mathrm{jet}\left(T\right) +
\xi_\mathrm{cjt}\left(T\right)$, we find:
\begin{align}\label{eqn:launchdate}
  T_\mathrm{launch} = T-\frac{\xi_\mathrm{tot}\left(T\right)}
                             {\xi_\mathrm{tot}\left(T+\Delta T\right)
                              -\xi_\mathrm{tot}\left(T\right)}
                        \,\Delta T \,.
\end{align}
The quantity $\xi_\mathrm{tot}$ has the advantage that it is just the angular
separation between the jet and counterjet bolides, and so can be determined
independently of the absolute location of the centre. This is particularly
useful for self-calibrated interferometric maps of SS~433, where the precise
core location can be hard to identify.

Similarly, we can define a total proper motion
\mbox{$\mu_\mathrm{tot}\left(T\right) := \mu_\mathrm{jet}\left(T\right) +
  \mu_\mathrm{cjt}\left(T\right)$} as the rate at which the two bolides move
apart. Using Equation \ref{eqn:propermotiondefinition}, this total proper
motion can be written:
\begin{align}\label{eqn:totalpropermotion}
  \mu_\mathrm{tot} = \frac{\xi_\mathrm{tot}\left(T\right)}{T-T_\mathrm{launch}}
                   = \frac{2c}{D}\frac{\beta_\perp}{1-\beta_\parallel^2}
\end{align}
where $\beta_\perp = \beta \sin{\theta}$.

We can rearrange Equation \ref{eqn:totalpropermotion} to find the
velocity component perpendicular to the line-of-sight, $\beta_\perp$. Then,
the jet speed itself can be written in terms of the total proper motion,
$\mu_\text{tot}$, and the line-of-sight velocity component, $\beta_\parallel$:
\begin{align}\label{eqn:jetspeed}
  \beta = \sqrt{\beta_\perp^2 + \beta_\parallel^2}
        = \sqrt{
          \left[\left({D}/{2c}\right)\mu_\mathrm{tot}
                \left(1-\beta_\parallel^2\right)\right]^2
          + \beta_\parallel^2
          }
\end{align}
where the line-of-sight velocity component, $\beta_\parallel$ is calculated
using Equation \ref{eqn:lineofsightvelocity}. 

\section{2011-12 VLBA observations}\label{sec:2011-12vlba}
\begin{figure*}
  \includegraphics[width=504.0pt]{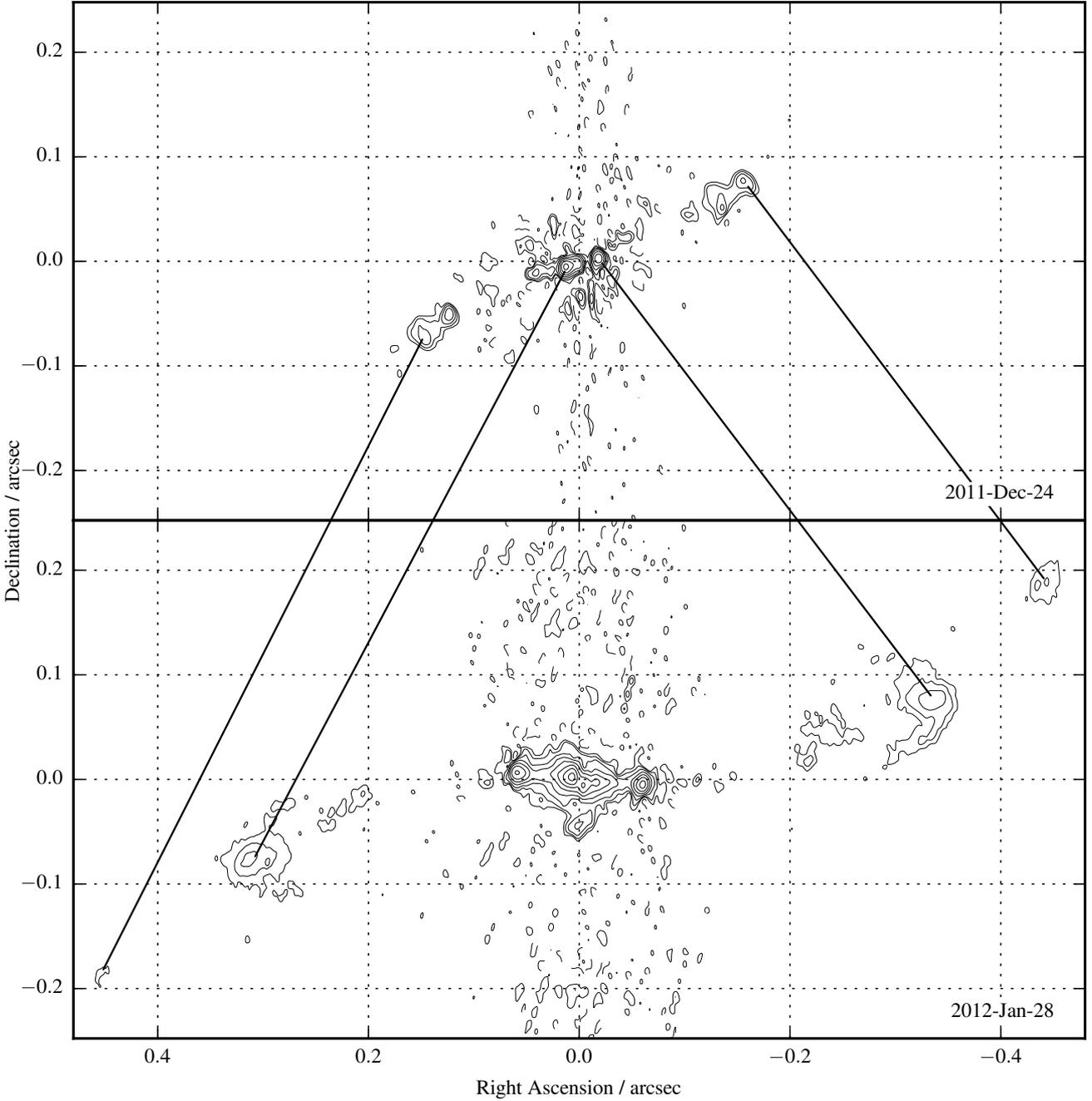}
  \caption{VLBA Observations of SS~433 in 2011-12. Contour levels are at $2^n$
    times the minimum contour level, which is
    $2.2\,\text{mJy}\,\text{beam}^{-1}$ in the upper panel, and
    $0.22\,\text{mJy}\,\text{beam}^{-1}$ in the lower panel. One negative
    contour (dashed) at ${-1}$ times the minimum contour level is also
    shown. The connecting lines identify the same ejections across the two
    epochs.}
  \label{fig:BCfig}
\end{figure*}
\begin{table*}
  \begin{tabular}{c  d{3.1} d{3.1} d{3.1}  d{3.1} d{3.1} d{3.1}  
    d{5.1} @{$\,\pm\,$} d{1.1} 
    d{1.3} @{$\,\pm\,$} d{1.3} 
    d{1.3} @{$\,\pm\,$} d{1.3}}
    \hline
    \multirow{2}{*}{Bolide}
    & \multicolumn{3}{c}{2011-Dec-24}
    & \multicolumn{3}{c}{2012-Jan-28}
    & \multicolumn{2}{c}{$T_\text{launch}$}
    & \multicolumn{2}{c}{Line-of-Sight}
    & \multicolumn{2}{c}{Bulk Speed} \\
    & \xi_\mathrm{jet} & \xi_\mathrm{cjt} & \xi_\mathrm{tot}
    & \xi_\mathrm{jet} & \xi_\mathrm{cjt} & \xi_\mathrm{tot}
    & \multicolumn{2}{c}{($\mathrm{MJD}$)} 
    & \multicolumn{2}{c}{Velocity $\beta_\parallel = v_\parallel / c$} 
    & \multicolumn{2}{c}{$\beta = v / c$} \\ \hline
    A & 162.2 & 173.9 & 336.1 & 488.9 & 480.5 & 969.4 &
       55901.3 & 0.9 &  0.009 & 0.029 & 0.288 & 0.017 \\
    B & 12.7  &  18.7 &  31.2 & 318.3 & 343.8 & 662.0 &
       55918.1 & 0.6 & -0.039 & 0.043 & 0.289 & 0.017 \\
    \hline
  \end{tabular}
  \caption{Angular displacements and derived launch dates, line-of-sight
    velocities, and launch speeds for bolides A and B in the 2011-12
    observations. Angular displacements are given in milliarcseconds from the
    core (located at the origin of Figure \protect\ref{fig:BCfig}). Speeds are
    given in units of $c$.}
  \label{tab:2011-12_bolidedata}
\end{table*}
SS~433 was observed with the Very Long Baseline Array (VLBA) on
$\rm{MJD}\,55919.83$ and $\rm{MJD}\,55954.75$ (i.e. on 2011-Dec-24, and 35
days later on 2012-Jan-28)\footnote{Modified Julian Date (MJD) is days since
  midnight on 1858-Nov-17. This is JD - 2400000.5.}. These images are shown in
Figure \ref{fig:BCfig}. These were \textit{L}-Band observations at
$1.6\,\text{GHz}$, giving a beam size of approximately $10\,\text{mas} \times
5\,\text{mas}$. The low frequency end of the band was grossly affected by
radio frequency interference (RFI) and flagged. This gave us $6$ contiguously
spaced IFs of $16\,\text{MHz}$. After \textit{a priori} amplitude and bandpass
calibration, and fringe fitting to calibrate phase, these IFs were averaged
down to a single channel, giving an image of the continuum for each epoch.

The data were flagged, calibrated, and imaged using standard routines in the
\textsc{aips} software package (\cite{2003ASSL..285..109G}), following the
recipes in Appendix C of the \textsc{aips} Cookbook
(\cite{national1990aips}). Each 11 hour observation consisted of alternating 7
minute scans of SS~433, and 2 minute scans of the phase calibrator J1929+0507
($4\degr$ from the science target). The source rises and sets at different
times at each antenna, and it was found that the images were improved by using
only the middle 8 hours of each epoch, when the array coverage is fullest. To
handle the complex, extended structure present in the jets, we found it to be
essential to use the multi-scale implementation of \textsc{clean} available
within the \textsc{aips} task \aips{IMAGR} (\cite{2009AJ....137.4718G}). We
found the greatest success when using 2 \textsc{clean} scales, at 1 and 5
times the beam size.

A significant challenge in imaging arises because of time variation in SS~433,
and especially from the motion of the jet bolides during the course of the
observations. The length of our observations give us sufficient
signal-to-noise to image the expanded jet bolides over an unparalleled period
after launch. But, with proper motions of $\sim 8 \,\text{mas}\,\text{d}^{-1}$
and observations lasting almost half a day, the bolides move by almost a full
beam in the course of the observations, violating the assumptions underlying
the use of the Fourier transform in synthesis imaging. This is the principal
cause of the spurious signals North and South of the main jet axis. Attempts
to mitigate this by imaging subsections of the full scan were only partially
successful. Nonetheless, the images we obtained are of sufficient quality for
us to determine the locations of the jet bolides securely. In each figure, the
lowest contour level is set at $\gtrsim 5\sigma$, where $\sigma$ is the rms
background noise away from the artefacts north and south of the
jet. Accordingly, we are able to confidently locate even the faint outer
bolides in the second epoch.

These two images show the milliarcsecond-scale jets of SS~433 in greater
detail and over greater extent than any previous VLBI image of the system. We
can trace the ejecta as they travel over $\gtrsim 0.4$ arcseconds on either
side of the core, or $\gtrsim 50\,\text{days}$ from their initial launch,
given that proper motions are of order of $8\,\text{mas}\,\text{d}^{-1}$. In
each epoch, we can identify a succession of pairs of bolides arising from
antiparallel ejections of knots of jet plasma. The time-spacing of the two
epochs is such that plasma that is launched shortly before the first
observation has not had time to fade before the second, revealing for the
first time the remarkable degree to which the jet knots expand as they fade
(the implications of which will be discussed in a forthcoming paper (Jeffrey
et al. in prep)).

For the two ejection complexes identified in both epochs, we can measure their
angular displacements from an estimated core position, located at the origin
in Figure \ref{fig:BCfig}. Then, we evaluate line-of-sight velocity,
launch date, and jet bulk speed using Equations \ref{eqn:lineofsightvelocity},
\ref{eqn:launchdate}, and \ref{eqn:jetspeed} respectively.

The resulting values are given in Table~\ref{tab:2011-12_bolidedata}. The two
ejections occur on $\mathrm{MJD}\,55901.3 \pm 0.9$ and $\mathrm{MJD}\,55918.1
\pm 0.6$ (i.e., $18.5\pm 0.9$ and $1.7\pm 0.6$ days before the first
observation). Both ejections launch bolides at speeds of almost $0.29c$. This
is the first unique measurement of the bulk speeds of the radio jets on
milliarcsecond scales. It is particularly noteworthy that the speeds lie at
the upper end of the range of those fitted to the arcsecond-scale radio jets
in \cite{2004ApJ...616L.159B}, and also that they match the high optical jet
speeds that \cite{2011MNRAS.417.2401B} reported as being associated with
flaring behaviour. We will discuss this further in Section
\ref{sec:discussion}.

\FloatBarrier
\section{Archival VLBA observations - the SS~433 movie}\label{sec:miodmovie}
We can also apply the methods derived in Section \ref{sec:calculations} to
historic VLBA observations of SS~433 in 2003.

SS~433 was observed on 39 out of the 42 days between 2003-June-26 and
2003-August-06, using 2 hour observation tracks with the VLBA at
$1.4\,\text{GHz}$ (\textit{L}-Band). The data were reduced by one of the
authors (AJM) with the \textsc{aips} software using standard techniques for
VLBI imaging (as discussed in the \textsc{aips} Cookbook). Multiscale
\textsc{clean} was used to assist with imaging, as was careful boxing, guided
by the predicted jet loci generated from the kinematic model. A movie of the
colour-scale images from this campaign has previously been released to the
community (\url{http://www.nrao.edu/pr/2004/ss433/};
\cite{2004nrao.pres....1.}). The individual observations are plotted as
contour maps in Figure ~\ref{fig:miodmovie}, and an animated version of these
contour plots is available with the online Journal.

From a qualitative inspection of Figure~\ref{fig:miodmovie}, we can see that
the jet consists mainly of a series of discrete ejection pairs, which we are
able to track over multiple successive days' observing as they travel away
from the core in an apparently ballistic fashion. In each case, both
components of a pair of bolides appear to be launched simultaneously. In the
first observation, 2 bolide pairs can be distinguished. Over the course of the
6 week campaign, a further 8 pairs are launched.

Throughout the campaign, the jet shows deviations from the kinematic model
predictions (i.e. the crosses in Figure~\ref{fig:miodmovie}, using the
parameter fits of \cite{2001ApJ...561.1027E}), implying variations in launch
angle or in jet speed (or in both) that are not described by this simple
model. \cite{2004AAS...20510401S} noted that the proper motions deviate from
those predicted by the model. In Section~\ref{subsec:propermotion2003} below,
we apply Equation~\ref{eqn:jetspeed} to show how these correspond to
variations in the physical jet launch speed.

It is also interesting to note that SS~433's behaviour over the course of the
2003 campaign appears to be divided into two distinct phases. During the first
part of the campaign, we observe a period of relative quiescence. Discrete jet
ejections can be identified, but they have low luminosities relative to later
bolides, and appear to lie within a more continuous background jet flow.

From about day 19 (i.e. from 2003-Jul-14, MJD 52834) until the end of the
observing campaign, the jet appears to move to a more active phase featuring
more intense bolide ejections. The jet appears more fragmented, with more
pronounced gaps in the jet flow between individual ejections (note that all
frames of Figure~\ref{fig:miodmovie} are contoured at the same levels). This
fragmentation suggests that the continuous jet seen in earlier epochs is
actually disrupted, although it may be a dynamic range or $uv$-plane sampling
effect.

We believe this second phase of behaviour to be the milliarcsecond-scale
manifestation of a radio flare, and in Section~\ref{sec:discussion} we will
discuss this hypothesis in relation to existing observations of SS~433.

\subsection{Proper motion analysis of the 2003 data} 
\label{subsec:propermotion2003} We can determine the bulk launch speeds
of the individual jet ejecta by using the analysis of Section
\ref{sec:calculations}. The peak positions of individual bolides were
determined using the \textsc{aips} task \aips{JMFIT}. At each epoch, a single
Gaussian component was fitted to each bolide; the peak positions of these
Gaussians served as the bolide locations. These fits are available in
machine-readable format with the online Journal.

\begin{landscape}
\begin{figure}
  \includegraphics[width=676pt]{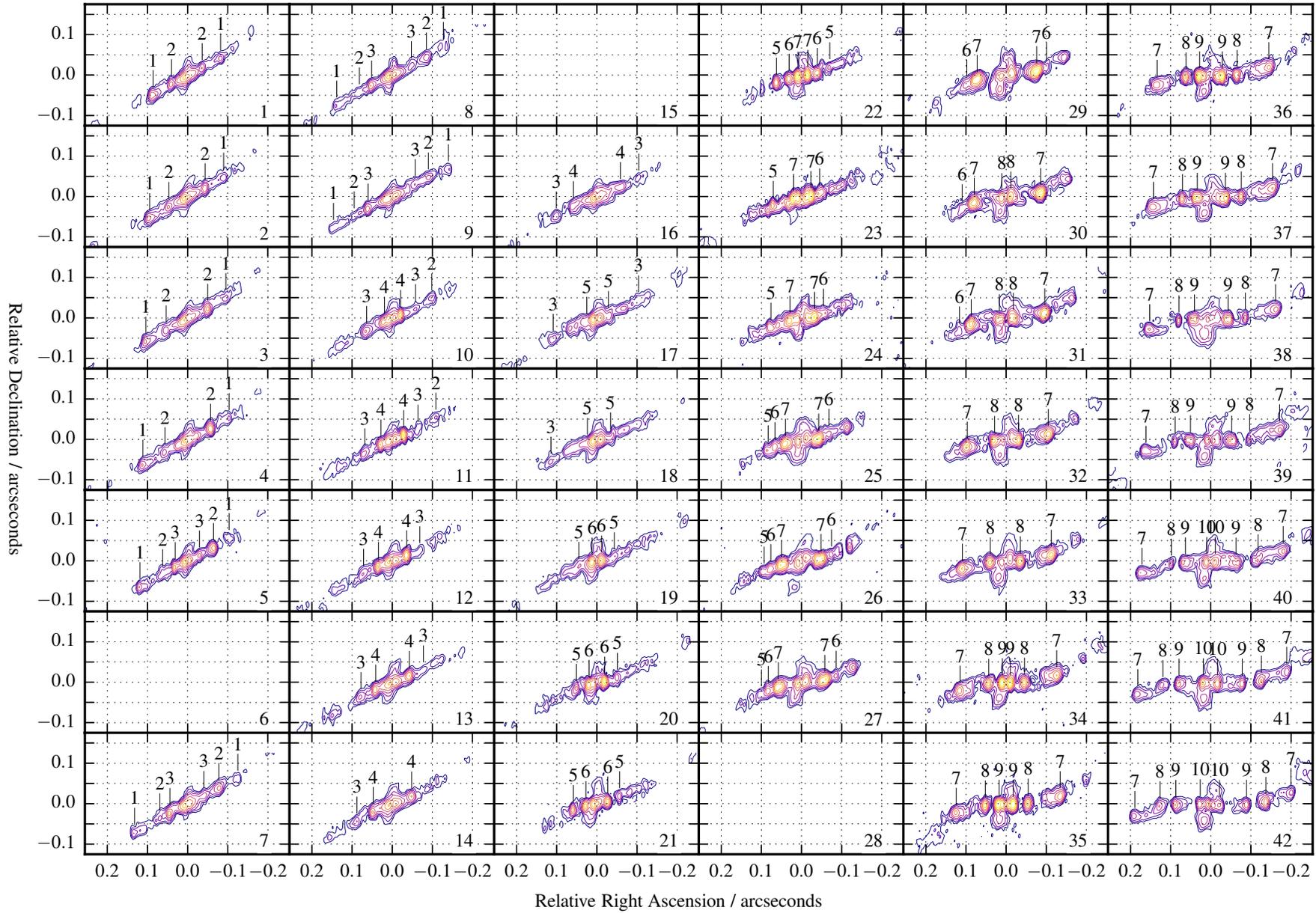}
  \caption{Contour Maps showing 2003 VLBA Observations of SS~433. Contour
    levels are identical across all frames, set at $2^n \times
    0.7\,\mathrm{mJy}\,\text{beam}^{-1}$ for integer $n$. The frame number
    corresponds to number of days from 2003-June-25 (i.e. $\rm{MJD} -
    52815.0$). The crosses denote the Kinematic Model predictions for bolides
    launched at 5 day intervals.}
  \label{fig:miodmovie}
\end{figure}
\end{landscape}

Next, we determine the proper motions, $\mu_\text{jet}$ and $\mu_\text{cjt}$,
of the bolides in the jet and counterjet. To do this, we fit the displacements
of the bolides from the launch point in each epoch assuming simultaneous and
oppositely directed launch, followed by ballistic propagation. In Figure
\ref{fig:miodmovie}, each frame has been centred to the estimated launch
point, taken to be the gap between the two stationary peaks that correspond to
the base of the jets. There are 4 parameters in this model: the proper motions
$\mu_\text{jet}$ and $\mu_\text{cjt}$, together with the position angle of
their trajectories on the sky, and the epoch of launch,
$T_\mathrm{launch}$. That is,
\begin{align*}
  \xi_{\alpha,\mathrm{jet}} &=  \mu_{\mathrm{jet}}\left(T-T_\mathrm{launch}
                                 \right)\cos{\chi}\sec{\delta_0} \\
  \xi_{\delta,\mathrm{jet}} &=  \mu_{\mathrm{jet}}\left(T-T_\mathrm{launch}
                                 \right)\sin{\chi} \\
  \xi_{\alpha,\mathrm{cjt}} &= -\mu_{\mathrm{cjt}}\left(T-T_\mathrm{launch}
                                 \right)\cos{\chi}\sec{\delta_0} \\
  \xi_{\delta,\mathrm{cjt}} &= -\mu_{\mathrm{cjt}}\left(T-T_\mathrm{launch}
                                 \right)\sin{\chi} 
\end{align*}
where $\xi_\alpha$ and $\xi_\delta$ denote shifts in Right Ascension and
Declination, the position angle, $\chi$, is the angle formed between the jet
and the E-W axis, and $\delta_0$ is the Declination of the source
(+4\degr58\arcmin57\farcs764). We use a least-squares fitting routine, making
the simplifiying assumption of uniform, unknown errors in the
coordinates. Uncertainties on the fitted parameters are estimated by
evaluating the Hessian matrix of second derivatives at the maximum of the
posterior PDF, and then using this to approximate the PDF with a multivariate
Gaussian. 

\begin{figure}
  \includegraphics[width=240.0pt]{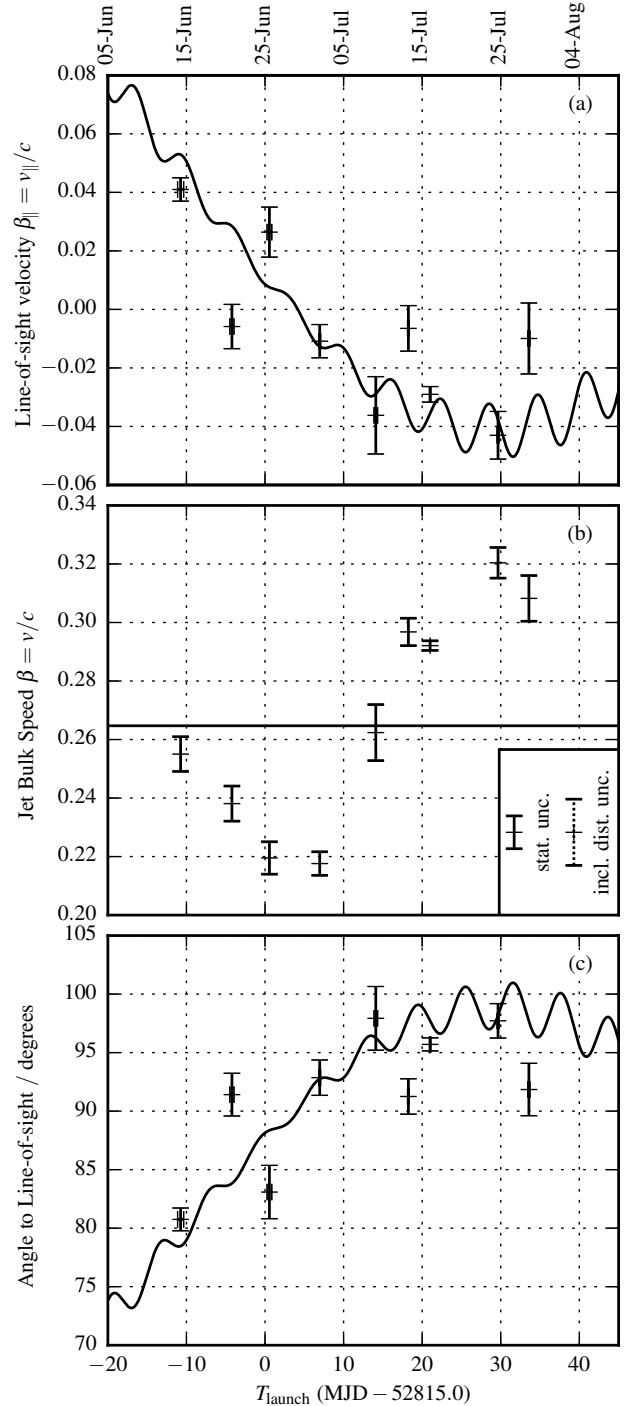}
    \caption{(a) Line-of-sight velocity component, $\beta_\parallel$, plotted
      against launch date. (b) Jet speed plotted against launch date. (c)
      Angle between jet bolide velocity vector and line-of-sight. In (a) and
      (c), the solid line shows the predicted values derived from the
      kinematic model, while the solid line in (b) shows the value of launch
      speed derived from optical spectroscopy by
      \protect\cite{2001ApJ...561.1027E}: $\beta = 0.2647 c$.  The systematic
      deviations from this constant value are clear. Uncertainties shown in
      (a) and (c) are $1\sigma$. The errorbars in (b) indicate only the
      uncertainty due to statistical variations in the fitted parameters. We
      have indicated the effect of the uncertainty in the distance to the
      object in the inset pane, where the mean statistical uncertainty in the
      fit (stat. unc.) and mean total uncertainty including that on the
      distance to SS~433 (incl. dist. unc.) are indicated. See
      Section~\protect\ref{sec:uncertainties}. The errors in launch date are
      smaller than the marker width on this scale.}
    \label{fig:miodmovie_jet_speed}
\end{figure}

The fitted proper motions are listed in Table~\ref{tab:miodmovie_results}
(there are too few data points for a reliable fit to the final ejection,
giving a total of 9 fitted bolide pairs).  Using equations
(\ref{eqn:lineofsightvelocity}) and (\ref{eqn:jetspeed}), the line-of-sight
velocity components and the bulk speeds were calculated for each ejecta
pair. They are plotted against epoch of launch in
Figure~\ref{fig:miodmovie_jet_speed}. The corresponding angles to the
line-of-sight are also plotted, showing the precession of the jet launch
vector. Uncertainties are estimated by drawing random samples from the
Gaussian approximation to the posterior PDF on the fitted parameters, together
with the distance to SS~433, \mbox{$D=5.5\pm 0.2 \,\text{kpc}$}
(\cite{2004ApJ...616L.159B}). When propagated through equations
(\ref{eqn:lineofsightvelocity}) and (\ref{eqn:jetspeed}), this gives estimates
for the errors on the derived parameters. In fact, the uncertainty on distance
$D$ is the largest source of uncertainty in the estimation of jet speed
$\beta$. This is shown in Figure~\ref{fig:miodmovie_jet_speed}(b); we will
return to this in Section~\ref{sec:uncertainties}.

Inspecting the derived speeds, we see quantitative evidence for a change in
behaviour during the campaign. Jet speeds are low ($\beta \lesssim 0.26$)
for ejections before approximately day 15. In the later observations, jet
speeds are high, exceeding $0.29c$. 

\begin{table*}
  \begin{tabular}{c 
    d{3.3} @{$\,\pm\,$} d{1.3} 
    d{2.3} @{$\,\pm\,$} d{1.3} 
    d{2.3} @{$\,\pm\,$} d{1.3} 
    d{2.3} @{$\,\pm\,$} d{1.3} 
    d{3.1} @{$\,\pm\,$} d{1.1}
    d{1.3} @{$\,\pm\,\left(\right.$} d{1.3}@{$\,+\,$}d{1.3}@{$\left.\right)$}}
    \hline
    Bolide & \multicolumn{2}{c}{$T_\text{launch}$} & 
    \multicolumn{2}{c}{$\mu_\text{jet} / \mathrm{mas}\,\mathrm{d}^{-1}$} & 
    \multicolumn{2}{c}{$\mu_\text{cjt} / \mathrm{mas}\,\mathrm{d}^{-1}$} &
    \multicolumn{2}{c}{$\beta_\parallel = v_\parallel / c$} & 
    \multicolumn{2}{c}{$\theta / \degr$} &
    \multicolumn{3}{c}{$\beta = v / c$}  \\ \hline
     1 & -10.740 & 0.385 &   8.262 & 0.198 &   7.611 & 0.184 &  
           0.041 & 0.004 &  80.8 & 1.0 &   0.255 & 0.006 & 0.005 \\
     2 &  -4.211 & 0.264 &   7.449 & 0.209 &   7.536 & 0.185 &  
          -0.006 & 0.008 &  91.4 & 1.8 &   0.238 & 0.006 & 0.004 \\
     3 &   0.561 & 0.301 &   7.046 & 0.182 &   6.683 & 0.192 &  
           0.026 & 0.009 &  83.1 & 2.3 &   0.220 & 0.005 & 0.004 \\
     4 &   6.962 & 0.118 &   6.768 & 0.131 &   6.916 & 0.133 &  
          -0.011 & 0.006 &  92.9 & 1.5 &   0.218 & 0.004 & 0.005 \\
     5 &  14.098 & 0.231 &   7.894 & 0.211 &   8.487 & 0.359 &  
          -0.036 & 0.013 &  97.9 & 2.7 &   0.262 & 0.010 & 0.004 \\
     6 &  18.247 & 0.112 &   9.279 & 0.141 &   9.400 & 0.183 &  
          -0.006 & 0.008 &  91.3 & 1.5 &   0.297 & 0.005 & 0.007 \\
     7 &  21.014 & 0.074 &   8.891 & 0.056 &   9.422 & 0.058 &  
          -0.029 & 0.003 &  95.7 & 0.6 &   0.292 & 0.002 & 0.009 \\
     8 &  29.654 & 0.126 &   9.582 & 0.178 &  10.443 & 0.189 &  
          -0.043 & 0.008 &  97.7 & 1.5 &   0.320 & 0.005 & 0.007 \\
     9 &  33.583 & 0.135 &   9.602 & 0.269 &   9.795 & 0.271 &  
          -0.010 & 0.012 &  91.8 & 2.2 &   0.308 & 0.008 & 0.006 \\
    \hline
  \end{tabular}
  \caption{Best fits to proper motions and bolide launch dates, together with
    derived line-of-sight velocities $\beta_\parallel$, angles $\theta$ to the
    line-of-sight, and bulk speeds $\beta$, for the first 9 bolides seen in
    the 2003 campaign. $T_\mathrm{launch}$ is given as $\mathrm{MJD} -
    52815.0$ (i.e. referred to 2003-June-25). The bolide numbers correspond to
    the labels in Figure~\protect\ref{fig:miodmovie}. Uncertainties are at the
    $1\sigma$ level. For the bulk speed, $\beta$, the error is given as
    statistical uncertainty in the fit $+$ systematic uncertainty in the
    distance $D$ to the source (see discussion in
    Section~\protect\ref{sec:uncertainties}).}
  \label{tab:miodmovie_results}
\end{table*}

\section{Discussion}\label{sec:discussion}

\subsection{Do jet speeds and brightnesses rise in flaring episodes?} 
\begin{figure}
  \includegraphics[width=240.0pt]{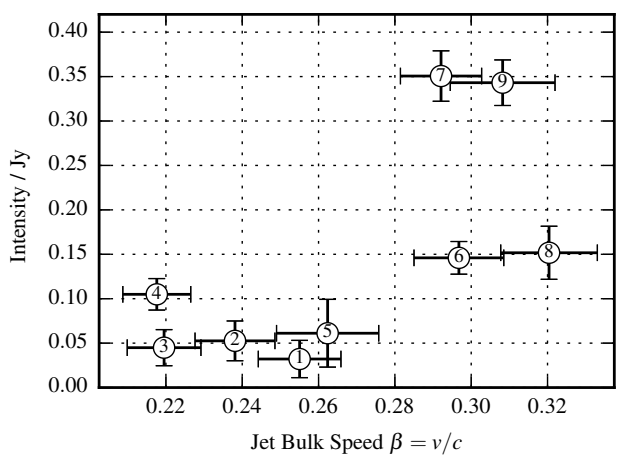}
    \caption{Jet launch speed plotted against intensity for the bolides seen
      in the 2003 campaign. Intensity is calculated as the peak of the mean of
      the deboosted intensities for the jet and counterjet. The numbers within
      the markers denote the bolide's number in the sequence of ejections
      2003-Jun-14 to 2003-Jul-28 (see Figure~\protect\ref{fig:miodmovie} and
      Table~\protect\ref{tab:miodmovie_results}).}
    \label{fig:speed-intensity_plot}
\end{figure}
\begin{table}
  \begin{center}
  \begin{tabular}{c 
    d{1.3} @{$\,\pm\,$} d{1.3} 
    d{1.3} @{$\,\pm\,$} d{1.3} 
    d{1.3} @{$\,\pm\,$} d{1.3}}
    \hline
    \multirow{2}{*}{Bolide}
    & \multicolumn{2}{c}{Bulk Speed} 
    & \multicolumn{4}{c}{Flux on $\mathrm{MJD}\,55919.83$ / $\mathrm{Jy}$} \\
    & \multicolumn{2}{c}{$\beta = v / c$}
    & \multicolumn{2}{c}{Jet} 
    & \multicolumn{2}{c}{Counterjet} \\ \hline
    A & 0.288 & 0.017 & 0.126 & 0.003 & 0.095 & 0.002 \\
    B & 0.289 & 0.017 & 0.518 & 0.018 & 0.226 & 0.008 \\
    \hline
  \end{tabular}
  \end{center}
  \caption{The jet bulk speeds and deboosted fluxes for the two bolide pairs
    seen in the $\rm{MJD}\,55919$ (2011-Dec-24) observation. The reported
    fluxes are integrated fluxes measured using \textsc{aips} task
    \aips{JMFIT}. Note that these fluxes are lower limits for the peak
    brightness these bolides reach, but that these bolides still obey the
    ``bright-fast'' pattern seen for other bolides launched during radio flares
    (c.f. Figure~\protect\ref{fig:speed-intensity_plot}).}
  \label{tab:fluxes2011}
\end{table}
The 2003 VLBA observations allow us to characterise the distinction between
the jets' behaviour in quiescent and flared periods of activity. From
Figure~\ref{fig:miodmovie_jet_speed}, we can clearly see the distinct speed
distributions before and after day~$15$: speeds do rise during this flaring
episode.

The 2003 VLBA observations make clear that more intensely radiating bolides
travel faster than fainter bolides. This is plotted in
Figure~\ref{fig:speed-intensity_plot}, where peak intensity is plotted against
the calculated jet speed for the 9 bolide pairs in the 2003 data whose
positions can be securely measured in more than 4 epochs. `Intensity' here is
a representative estimate; it is found by deboosting the total intensity
within integrated fitted Gaussians returned by \aips{JMFIT}, averaging over
jet and counterjet ejections at each epoch, and then selecting the epoch for
which this has its maximum value. For details on extracting the true deboosted
brightness from a relativistic synchrotron source, see
\cite{2004ApJ...603L..21M}; here, we have assumed optically thin synchroton
emission from a jet composed of discrete plasmons, adiabatically expanding
with a constant rate of change of radius, and with electron energy spectrum
$\gamma^{-2.2}$, where $\gamma$ is the electron Lorentz factor (not the bulk
Lorentz factor of the jet bolides). It is, however, worth remembering that for
SS~433, whose jets are only mildly relativistic and lie close to the plane of
the sky, Doppler boosting is a small correction (a few percent, rather than
orders of magnitude).

The two 2011 bolides also fit this pattern (Table~\ref{tab:fluxes2011}). They
are not plotted in Figure~\ref{fig:speed-intensity_plot} because it is not
possible to determine their peak brightnesses (since we do not have
measurements of the evolution of their lightcurves). However, we can take the
averaged and deboosted fluxes from the first of the two observations
($\rm{MJD}\,55919$, i.e., 2011-Dec-24) observation as lower limits. These
bolides' properties (averaged flux densities $0.11$ and $0.37\,\mathrm{Jy}$
for bolide pairs A and B, with speeds $0.288c$ and $0.289c$ respectively) are
consistent with the ``bright-fast'', flare characteristics displayed by
bolides 6, 7, 8, and 9 during the 2003 campaign.

So, the bolides from both the 2003 and 2011 campaigns display a clear trend:
\emph{`fast' ejections (those exceeding $0.28c$) can be as much as 8 times as
  luminous as those `slow' ejections with speeds below $0.26c$}.

Note that not all fast ejections are especially luminous (c.f. bolides 6 and
8). However, in Figure~\ref{fig:speed-intensity_plot}, we can again see the
separation of the ejections in flaring episodes (bolides 6-9), and those in
quiescence (bolides 1-5). This reinforces the conclusion that flaring
episodes are characterised by generally faster and more luminous ejections,
and that the fastest, most luminous ejections only occur during these periods
of elevated activity.

This is entirely consistent with the conclusion of \cite{2011MNRAS.417.2401B}:
high jet speeds as inferred from the \emph{optical} emission lines (which are
themselves more intense than usual) precede radio flares. Now, by using only
radio maps to infer jet speeds completely independently of optical spectra, we
also know that radio flares are characterised by high speeds in the
\emph{radio} jets. This further reinforces the link between the optical and
radio ejecta.

\subsection{How do these data compare with known properties of flares?}
\begin{figure}
  \includegraphics[width=240.0pt]{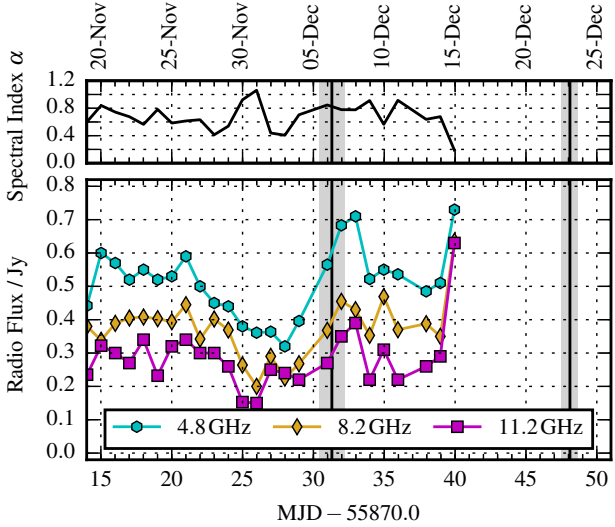}
    \caption{Bottom Panel: Daily radio-frequency fluxes for the period of the
      2011 observations, as measured by the RATAN telescope. Top Panel:
      Corresponding spectral indices, $\alpha$, for the same period (note that
      flux \mbox{$S \propto \nu^{-\alpha}$}). The black vertical lines denote
      the launch dates of the two ejections A and B as calculated in
      Section~\protect\ref{sec:2011-12vlba}, with the associated uncertainties
      in grey.}
    \label{fig:ratanradio}
\end{figure}
Our understanding of this phenomenon has been guided by our association of the
behaviour seen on milliarcsecond scales in these maps with the flaring
episodes seen in optical spectra and in photometric radio data. In the case of
the 2011 data, we can make this comparison slightly more explicit.

It is unfortunate that there are not contemporaneous optical monitoring data
for this object covering the periods of these two VLBA campaigns. However, for
the 2011 observations, there is monitoring data from the RATAN telescope for
the period up to 2011-Dec-15 (\cite{2003BSAO...56...57T}). This overlaps with
the launch of bolide pair A on 2011-Dec-06. In the total intensity radio data
(Figure~\ref{fig:ratanradio} lower panel), there is evidence that the radio
intensity is quenched from 2011-Nov-26, and then rises by perhaps a factor of
$3$ over the $5$ or so days preceding the ejection of bolide pair A (marked by
the first grey bar). This pattern fits with the Type II flare described in
\cite{2011MNRAS.417.2401B}, where Type II radio flares were associated with
the launch of optical jets. This is the first time that we have seen evidence
in SS~433 that a Type II radio flare corresponds to the launch of a
radio-emitting jet bolide pair.

As an aside, we emphasise that not all radio flares seen in photometric
monitoring are associated with the jets. Type I flares in
\cite{2011MNRAS.417.2401B} are characterised by a flat spectrum, and are
associated with radio emission from an enhanced wind off the accretion
disc. The peak in radio intensity at the end of the RATAN monitoring data on
2011-Dec-15 displays a flat radio spectrum, and does not appear to be
associated with a bolide launch. 2011-Dec-15 is midway between the launches of
bolides A and B -- a bolide launched on this date would appear in the gap
between these bolides in Figure~\ref{fig:BCfig}, unless its initial brightness
was so faint that it had faded completely before the observation on
2011-Dec-24 (which seems unlikely, given that the peaks in the RATAN data on
days 33 and 40 in Figure~\ref{fig:ratanradio} are of the same
size). Consequently, we believe this peak shows the hallmarks of a
disc-wind-based radio flare, preceding the launch of a large jet ejection on
2011-Dec-23.

\subsection{The circumbinary ruff during radio flares}
Several previous VLBI observations of SS~433 have revealed the presence of a
circumbinary ruff -- an extended region of radio emission aligned above and
below the core on the sky, and not associated with the jet (e.g.,
\cite{1999A&A...348..910P},
\cite{2001ApJ...562L..79B}). \cite{2009ApJ...698L..23D} explored the
precession of this ruff using all existing VLBI data -- including the
observations from 2003 that are shown here in Figure~\ref{fig:miodmovie}. They
suggested that the circumbinary ruff is a large scale manifestation of an
outflow via a quasi-stable circumbinary disc. This circumbinary disc exists on
scales comparable to a few times the orbital separation, and has been observed
as static optical emission lines by \cite{2008ApJ...678L..47B}).

We comment that, during the 2003 observations, the ruff appears to grow
significantly in size as the flare develops after day 19. We suggest that this
may arise from one or more of the following: a) the observed variation is an
artefact of changes in $uv$-coverage; b) the ruff appears larger because it is
illuminated by more radiation from a brighter, flaring core; or c) the ruff is
larger because of enhanced mass outflow from the circumbinary disc. In the
first case, we note the apparent dimming of the ruff on days 23 and 26 is
suggestive that the effect may be partly artefactual. We also note that the
ruff's extent of $\gtrsim 50\,\text{mas}$ is at the upper limit of angular
scales to which the VLBA is sensitive, because the shortest VLBA baselines
give only sparse coverage of the centre of the $uv$-plane. The third case is
intriguing given optical observations of \cite{2011MNRAS.417.2401B} that
provided evidence that the inner circumbinary disc structure was disrupted
during a flare in 2004-October. In the 2003-July observations, there are hints
of detatchment between the core components and the radio ruff from day 31
onwards, raising the possibility that the enlarged ruff is a manifestation of
an ejected circumbinary disc. However, this would require a very fast outflow
speed -- at $5.5\,\text{kpc}$, $50\,\text{mas}$ corresponds to a physical size
of $4 \times 10^{13} \,\text{m}$. To cover this in $\sim 10\,\text{d}$ would
require an outflow speed exceeding $40,000\,\text{km}\,\text{s}^{-1}$, which
seems improbably high (c.f. \cite{2004nrao.pres....1.}).

\subsection{The unified picture}
Despite the lack of contemporaneous optical and radio monitoring data, there
are enough similarities between what we see in these milliarcsecond scale
observations and what is seen at other scales and frequencies that we can draw
out some general characteristics.

In Figure~\ref{fig:miodmovie_jet_speed}, we see that jet launch speeds appear
to fall over a period of about $30$ days before the flare begins and jet
speeds increase dramatically on or around day 15. A similar decline in speed
is seen in the optical jet lines in figure 6(e) of \cite{2011MNRAS.417.2401B},
where jet speeds decline from $\sim 0.27c$ to $\sim 0.24c$ over the 20 days
prior to a flare.

Both in the 2011-12 observations, and during the latter half of the 2003
campaign, we see bright, fast ejections, punctuated by gaps of 5 to 10
days. Again, almost identical behaviour is seen by \cite{2011MNRAS.417.2401B}
in the optical jet lines, where the intensity of the optical jet lines is
lowest in the days immediately preceding the start of the flare. This stands
in contrast to the behaviour of the jet in the early days of the 2003
campaign, where the jet has a more steady character. The difference is most
clearly seen by comparing the first and last frame of
Figure~\ref{fig:miodmovie} (noting that the images of all epochs are contoured
to the same minimum level): in the first frame we see faint bolides set
against an apparently steady background flow, while in the last we see a
succession of bright, compact bolides separated by gaps of low/no emission.

Taken together, we note that the variations in both the speeds of the optical
emission lines and the speeds determined by the milliarcsecond scale bolides
possess remarkably similar characteristics. In the absence of simultaneous,
time-resolved optical and radio observations, this coincidence is intriguing
evidence that the ejecta revealed by optical emission are indeed directly
related to the ejecta seen through radio emission. That is, that both the
optical emission lines and the radio continuum emission originate from the
same jet moving ballistically and seemingly with the same speed, although this
expectation remains to be tested with quasi-simultaneous radio imaging and
optical spectroscopy.

\subsection{The effect of uncertainty in the distance $D$}
\label{sec:uncertainties} As noted in Section~\ref{sec:miodmovie}, the
dominant contribution to the uncertainty in the calculations in
Section~\ref{sec:calculations} is the uncertainty in the distance $D$
(c.f. \cite{2003MNRAS.340.1353F}). For SS~433, we are fortunate to have a well
measured distance value, confirmed by two entirely different methods. Firstly,
fitting the precession model to the arcsecond-scale radio jets yields a
distance of $5.5\pm 0.2\,\text{kpc}$ (\cite{2004ApJ...616L.159B}). This
distance is consistent with the lower limit imposed on the distance by
sensitive VLA measurements of HI in absorption at a line-of-sight speed of
$75\,\text{km}\,\text{s}^{-1}$; absorption at this speed precludes distances
for SS433 below $5.5\,\text{kpc}$ (\cite{2007MNRAS.381..881L}), assuming the
validity of Galactic gas rotation models
(\cite{1983ApJS...53..591D}). Moreover measurements of HI in emission,
presented by \cite{2007MNRAS.381..881L} using the Green Bank Telescope, reveal
interactions of gas at the same rotation speed (i.e. the same distance) with
the W50 nebula that encloses SS~433. Most importantly, since this distance is
inferred from HI and Galactic rotation curves, it is totally independent of
any assumptions about SS~433's jet speeds; therefore our jet speed
determinations on milliarcsecond-scales are not prejudiced by fits to the jet
speeds on arcsecond scales. This gives us security in using the distance of
\cite{2004ApJ...616L.159B}.

In the inset panel to Figure~\ref{fig:miodmovie_jet_speed}(b) (and in
Table~\ref{tab:miodmovie_results}), we have indicated the effect of the
uncertainty in this distance $D$. The solid error bars assigned to each data
point in the plot show the uncertainties only due to the statistical
uncertainty (stat. unc.) in the fits to the proper motions
$\mu_{\mathrm{app}}$ and $\mu_{\mathrm{rec}}$. The dotted error bar in the
inset shows the scale of the uncertainties on each speed data point including
the distance uncertainty (incl. dist. unc.), i.e., when the distance $D$ is
allowed to be drawn from within the range $5.5\pm0.2\,\text{kpc}$. Since for
all measurements, the object must lie at the same distance, the effect of this
distance uncertainty is to give a systematic shift in the speed-axis values --
the statistical error bars show the uncertainties on each point relative to
one another, and give a truer reflection of our confidence that the jet
bolides observed in the 2003 campaign show a significant variation in their
launch speeds.

The essential point here is that an error in distance measurement would
manifest itself as a systematic error in the derived absolute launch speeds
across all ejections. It would \emph{not} change our most important overall
conclusion: faster jet launch speeds occur during flaring episodes.

\section{Conclusions}
The two campaigns of VLBA observations have given us unique measurements of SS
433's radio jet speeds on milliarcsecond scales, and shown that, like their
optical counterparts described in \cite{2011MNRAS.417.2401B}, flaring episodes
are associated with faster launch speeds in the radio jets. The 2003 campaign
indicates that there may be two quite different types of jet behaviour, namely
a flaring mode with a fast, bright jet consisting of compact, discrete
ejections, and a quiescent jet consisting of slower bolides set within an
apparently more continous background ejection.

Once again, high resolution observations of SS~433 have provided insights into
the remarkable complexity of the physics governing the behaviour of
microquasars. Contemporaneous observations in optical, radio and X-ray bands
are needed to elucidate this further.

\section*{Acknowledgements}
The data for the VLBA movie were reduced by AJM in 2003, and we are pleased to
thank Michael Rupen, Craig Walker, and Greg Taylor for helpful discussions
about VLBI imaging techniques in relation to these data.

RMJ thanks the Science and Technology Facilities Council for a PhD
studentship. This work arose in part from discussions during RMJ's visits to
Brandeis University and to NRAO in Socorro and Charlottesville, supported by
Merton College, Oxford, and by the Steve Rawlings Memorial Prize. We
particularly thank John Wardle, David Roberts, Tony Beasley, Alison Peck, and
Rick Perley for sharing their wisdom and insights. We are indebted to Eric
Greisen for his help with the finer points of \textsc{aips}.

The VLBA is a facility of the National Radio Astronomy Observatory, operated
by Associated Universities, Inc., under a cooperative agreement with the
National Science Foundation. The figures were produced using the Matplotlib
plotting package (\cite{Hunter:2007}). 



\bibliographystyle{mnras}
\bibliography{jetspeedspaper.bib}


\bsp	
\label{lastpage}
\end{document}